\documentclass[jcp,twocolumn,superscriptaddress,floatfix,graphicx,showpacs]{revtex4-1}
\usepackage{ifpdf}
 \newif\ifpdf
\ifx\pdfoutput\undefined
   \pdffalse
\else
   \pdfoutput=1
   \pdftrue
\fi
\ifpdf
   \usepackage{graphicx}
   \usepackage{epstopdf}
   \usepackage{color}
\else
   \usepackage{graphicx}
\fi

\usepackage{srctex}
\usepackage{hyperref}

\usepackage{amsmath,amssymb}

\usepackage{epsfig}

\usepackage{epsfig}

\begin{document}

\title{Nucleation instability in super-cooled Cu-Zr-Al glass-forming liquids}

\author{R.E. Ryltsev}
\affiliation{Institute of Metallurgy, Ural Branch of Russian Academy of Sciences, 620016, 101 Amundsena str., Ekaterinburg, Russia}
\affiliation{Ural Federal University, 620002, 19 Mira str., Ekaterinburg, Russia}
\affiliation{L.D. Landau Institute for Theoretical Physics, Russian Academy of Sciences, 119334, 2 Kosygina str., Moscow, Russia}
\author{B.A. Klumov}
%\affiliation{Aix-Marseille-Universit\'{e}, CNRS, Laboratoire PIIM, UMR 7345, 13397 Marseille cedex 20, France}
\affiliation{High Temperature Institute, Russian Academy of Sciences, 125412, 13/2 Izhorskaya str., Moscow, Russia}
\affiliation{L.D. Landau Institute for Theoretical Physics, Russian Academy of Sciences, 119334, 2 Kosygina str., Moscow, Russia}
\affiliation{Ural Federal University, 620002, 19 Mira str., Ekaterinburg, Russia}

\author{N.M. Chtchelkatchev}
%\affiliation{Department of Physics and Astronomy, California State University Northridge, Northridge, CA 91330, USA}
\affiliation{L.D. Landau Institute for Theoretical Physics, Russian Academy of Sciences, 119334, 2 Kosygina str., Moscow, Russia}
\affiliation{Moscow Institute of Physics and Technology, 141700б 9 Institutskiy per., Dolgoprudny, Moscow Region, Russia}
\affiliation{Institute for High Pressure Physics, Russian Academy of Sciences, 142190 Troitsk, Russia}
\affiliation{Institute of Metallurgy, Ural Branch of Russian Academy of Sciences, 620016, 101 Amundsena str., Ekaterinburg, Russia}
\affiliation{Ural Federal University, 620002, 19 Mira str., Ekaterinburg, Russia}

\author{K.Yu. Shunyaev}
\affiliation{Institute of Metallurgy, Ural Branch of Russian Academy of Sciences, 620016, 101 Amundsena str., Ekaterinburg, Russia}
\affiliation{Ural Federal University, 620002, 19 Mira str., Ekaterinburg, Russia}

\begin{abstract}
Special role in computer simulations of supercooled liquid and glasses is played by few general models representing certain classes of real glass-forming systems. Recently, it was shown that one of the most widely used  model glassformers --  Kob-Andersen binary Lennard-Jones mixture -- crystalizes in quite lengthy molecular dynamics simulations and, moreover, it is in fact a very poor glassformer at large system sizes. Thus, our understanding of crystallization stability of model glassformers is far from complete due to the fact that relatively small system sizes and short timescales have been considered so far. Here we address this issue for two embedded atom models intensively used last years in numerical studies of Cu-Zr-(Al) bulk metallic glasses. We consider ${\rm Cu_{64.5}Zr_{35.5}}$ and ${\rm Cu_{46}Zr_{46}Al_{8}}$ alloys as those having high glass-forming ability. Exploring their structural evolution at continuous cooling and isothermal annealing, we observe that both systems nucleate in sufficiently lengthy simulations, though ${\rm Cu_{46}Zr_{46}Al_{8}}$ demonstrate order of magnitude higher critical nucleation time. Moreover, ${\rm Cu_{64.5}Zr_{35.5}}$ is actually unstable to crystallization for large system sizes ($N > 20,000$). Both systems crystallize with the formation of tetrahedrally close packed Laves phases of different types. We reveal that structure of both systems in liquid and glassy state contains comparable amount of polytetrahedral clusters. We argue that nucleation instability of simulated ${\rm Cu_{64.5}Zr_{35.5}}$ alloy is due to the fact that its composition is very close to that for stable ${\rm Cu_2 Zr}$ compound with C15 Laves phase structure.
\end{abstract}

\maketitle
\section{Introduction}

Structural glasses, especially bulk metallic glasses (BMG), and supercooled liquids are in the shortlist of objects for study in modern chemistry and material science~\cite{Suryanarayana2017BMG_Book,Royall2015PhysRep,Xu2017JCP,Ediger2017JCP,Li2017JCP,Hu2017JCP,Wisitsorasak2017PNAS,Ninarello2017PRX,Ward2018ActaMater}. The reason is twofold: firstly, glasses have unique properties useful in industry, such as, corrosion
resistance~\cite{Frankel2018MaterDegr,Sagasti2018AIPAdv}, high specific strength \cite{Suryanarayana2017BMG_Book}, good thermoplastic formability \cite{Gong2018Intermet}, excellent bio compatibility \cite{Liens2018Materials} and, secondly, many aspects of structural glass formation still need an explanation \cite{Suryanarayana2017BMG_Book,Royall2015PhysRep,Cheng2011ProgMateSci}. Classical molecular dynamics (MD) has become the main theoretical tool that allows investigating properties of supercooled liquids and glasses which are hardly available in experiments. The examples include short-range and medium-range order ~\cite{Royall2015PhysRep,Cheng2011ProgMateSci,Pasturel2017JCP,Ghaemi2018JNCS}, dynamical heterogeneities \cite{Berthier2011RevModPhys,Usui2018JCP,Caballero2018JCP,Royall2017JStatMech}, potential energy landscape~\cite{Debenedetti2001Nature,Cavagna2009PhysRep,Handle2018JCP,Sun2017JCP} and nucleation mechanisms \cite{Shibuta2017NatureComm,Mokshin2017PCCP}.
The key point of any MD simulation is the choice of interaction potential determining all the system properties at microscopic level. A number of MD models have been thought for a long time to be ``representative fundamental'' glassformers building the foundation of the modern theory of structural glasses. Examples include Kob-Andersen~\cite{Kob1994PRL} and Wahnstr\"om~\cite{Wahnstrom1991PRA} binary Lennard- Jones mixtures representing simple liquids with pair isotropic interaction, and Beest-Kramer-van-Santen potential~\cite{Beest1955PRL} for silica, which is representative network-forming strong glassformer. However, recently, quite lengthy and accurate MD simulations have revealed  that some of these model glassformers are in fact unstable to crystallization. Instructive examples are Kob-Andersen and Wahnstr\"om models which were partially crystallized in lengthy MD~\cite{Toxvaerd2009JCP,Ingebrigtsen2018crystallisation}. Moreover, recent study reveals that Kob-Andersen mixture is in fact very poor glassformer for large ($N>10,000$ particles) system sizes~\cite{Ingebrigtsen2018crystallisation}.

Thus, our understanding of crystallization stability of model glassformers is far from complete due to the fact that relatively small system sizes and short timescales have been considered so far. Here we address this issue for two embedded atom models (EAM) developed in Refs.~\cite{Mendelev2009PhilMag,Cheng2009PRL}, which have been intensively used last years in numerical studies of Cu-Zr-(Al) bulk metallic glasses \cite{Li2009PRB,Cheng2009PRL,Peng2010ApplPhysLett,Soklaski2013PRB,Wu2013PRB,Wen2013JNonCrystSol,Wang2015JPhysChemA,Zhang2012JAlloysComp1,Ryltsev2016JCP,Klumov2016JETPLett,Khusnutdinoff2016JETP,Lad2017JCP}.

Among other glass-forming metallic alloys the Cu-Zr-based ones are of the greatest importance due to their high glass-forming ability~\cite{Xu2004ActMat,Wang2004AppPhysLett,Fan2017Intermet,Zhang2017JNCS}, relatively simple and cheap fabrication and possibility to form  high-strength ductile BMG composites \cite{Wu2011ActaMater,Hofmann2010Science,Rashidi2018MatSciEngA}. Besides, simulated Cu-Zr-based systems are widely accepted as models demonstrating pronounced icosahedral local ordering \cite{Li2009PRB,Cheng2009PRL,Peng2010ApplPhysLett,Soklaski2013PRB,Wu2013PRB,Wen2013JNonCrystSol,Wang2015JPhysChemA,Ryltsev2016JCP}.

We consider ${\rm Cu_{64.5}Zr_{35.5}}$ and ${\rm Cu_{46}Zr_{46}Al_{8}}$ alloys as those having high glass-forming ability. Exploring  their structural evolution at continuous cooling and isothermal annealing near the glass-transition temperature, we observe that, at sufficiently long times and large system sizes, both alloys nucleate with the formation of tetrahedrally close packed $\textrm{Cu}_2\textrm{Zr}$ Laves phases of different types. Thus, ${\rm Cu_{64.5}Zr_{35.5}}$ alloy crystallizes into the mixture of ${\rm Mg Cu_2}$ (C15) and ${\rm Mg Zn_2}$ (C14) phases while ${\rm Cu_{46}Zr_{46}Al_{8}}$ alloy crystallizes into the mixture of ${\rm Mg Cu_2}$ (C15) and ${\rm Mg Ni_2}$ (C36) phases. We also observe that ${\rm Cu_{64.5}Zr_{35.5}}$ is actually unstable to crystallization for large system sizes ($N > 20,000$). At the intermediate time scales where the simulated systems are still completely amorphous, their structure demonstrates pronounced polytetrahedral local order presented by Kasper polyhedra including icosahedra.

\section{Methods}

%Classical molecular dynamics (MD) is the main theoretical tool to study properties of glasses because it makes it possible to overcome the problems of analytical description of non-ordered condensed matter systems \cite{Liu1987PRB,Dubinin2014RussianChemRev,Dubinin2007JNonCrystSol,Dubinin2014JNonCrystSol} and allows studying microscopic structure and dynamics covering sufficiently large time and spatial scales.

For MD simulations, we use $\rm{LAMMPS}$ Molecular Dynamics Simulator \cite{Plimpton1995JCompPhys}. Periodic boundary conditions in Nose-Hoover NPT ensemble at $P=0$ were imposed. The number of particles varies from $N=5,000$ up to $N=50,000$.  The MD time step was 2 fs that provides good energy conservation at given thermodynamic conditions.

As the model of interaction between components of ${\rm Cu_{64.5}Zr_{35.5}}$ alloy we use widely accepted EAM potential developed by Mendelev et al. \cite{Mendelev2009PhilMag,Zhang2015PRB}. This potential has been specially designed to describe liquid and glassy states of the Cu-Zr alloys. For the ${\rm Cu_{46}Zr_{46}Al_{8}}$ alloy, we apply potential developed by Cheng et al. \cite{Cheng2009PRL}.

Initial configurations were prepared as hcp-lattice with random seeding of the species in the lattice sites. These configurations were melted, completely equilibrated at $T=1800$~K and then cooled down to $T=300$~K with different cooling rates $\gamma=\Delta T/\Delta t$ in the interval of $\gamma\in(10^{9}, 10^{13})$ K/s. A few configurations, collected from continuous cooling simulations at $\gamma=10^{9}$ and $\gamma=10^{10}$ K/s, were isothermally annealed for 0.5-3 $\mu$s at different temperatures in the vicinity of the glass transition temperature.

Note that such calculations require a fair amount of computer resources. For example, cooling a 20,000 particle system from 1800 K down to 300 K at cooling rate of $\gamma = 10^{9}$ K/s takes about a month of calculations on 128 computer cores in parallel.

To study the structure of both liquid, glassy and crystalline phases we use radial distribution functions $g(r)$, bond orientational order order parameters (BOOP) $q_l$~\cite{Steinhardt1981PRL,Steinhardt1983PRB,Hirata2013Science}, Voronoi tessellation (VT) \cite{Finney1977Nature,Brostow1998PRB}, diffraction analysis and visual analysis of the snapshots. Detailed description of these methods is presented in Ref.~\cite{Ryltsev2015SoftMatt,Ryltsev2016JCP}. Note that we use radical VT \cite{Gellatly1982JNCS} as implemented in the code developed by the W. Windl group at Ohio State University \cite{Ward2013PRB}. To create radical VT, the atomic radii for copper, zirconium and aluminum were determined from the locations of fist peaks of partial radial distribution functions. Thereafter, faces smaller than 1 \% of
the total surface area of each Voronoy polyhedron were removed in order to reduce the impact of thermal fluctuations on the topological structure of Voronoi polyhedra \cite{Troadec1998EPL,Brostow1998PRB,vanMeel2012JCP}.

As we will see below, the main structural motifs of the systems under consideration are Z10-Z16 Kasper polyhedra including both icosahedron Z12 and Z16 polyhedron as building blocks of Laves phases. Let us consider in more detail the criteria we use to identify these local clusters. According to VT method, the symmetry of a cluster (Voronoi polyhedron) is characterized by Voronoi index $\langle n_3, n_4, n_5, n_6\rangle$, where $n_i$ is the number of $i$-edged faces. The Z10-Z16 polyhedra under consideration have following Voronoi indexes \cite{Cheng2011ProgMateSci}: $Z10 \equiv\langle 0, 2, 8, 0 \rangle$,  $Z11 \equiv\langle 0, 2, 8, 1 \rangle$,  $Z12 \equiv\langle 0, 0, 12, 0 \rangle$, $Z13 \equiv\langle 0, 1, 10, 2 \rangle$,  $Z14 \equiv\langle 0, 0, 12, 2 \rangle$, $Z15 \equiv\langle 0, 0, 12, 3 \rangle$  and $Z16 \equiv\langle 0, 0, 12, 4 \rangle$.

For the sake of comparison with VT,  we also use BOOP to identify icosahedra as the most popular clusters.  Within the framework of BOOP, a cluster is treated as icosahedral-like one if it has BOO parameters $q_6 > 0.6$ and $w_6 < -0.16$ (for ideal non-disturbed icosahedron the values of such parameters are $q_6^{\rm (id)} = 0.663$ and $w_6^{\rm (id)} = -0.169$ ). Similar criteria can be determined for any cluster.

\section{Results}

\subsection{Structural evolution at continuous cooling}

 \begin{figure}
  \centering
  \includegraphics[width=0.8\columnwidth]{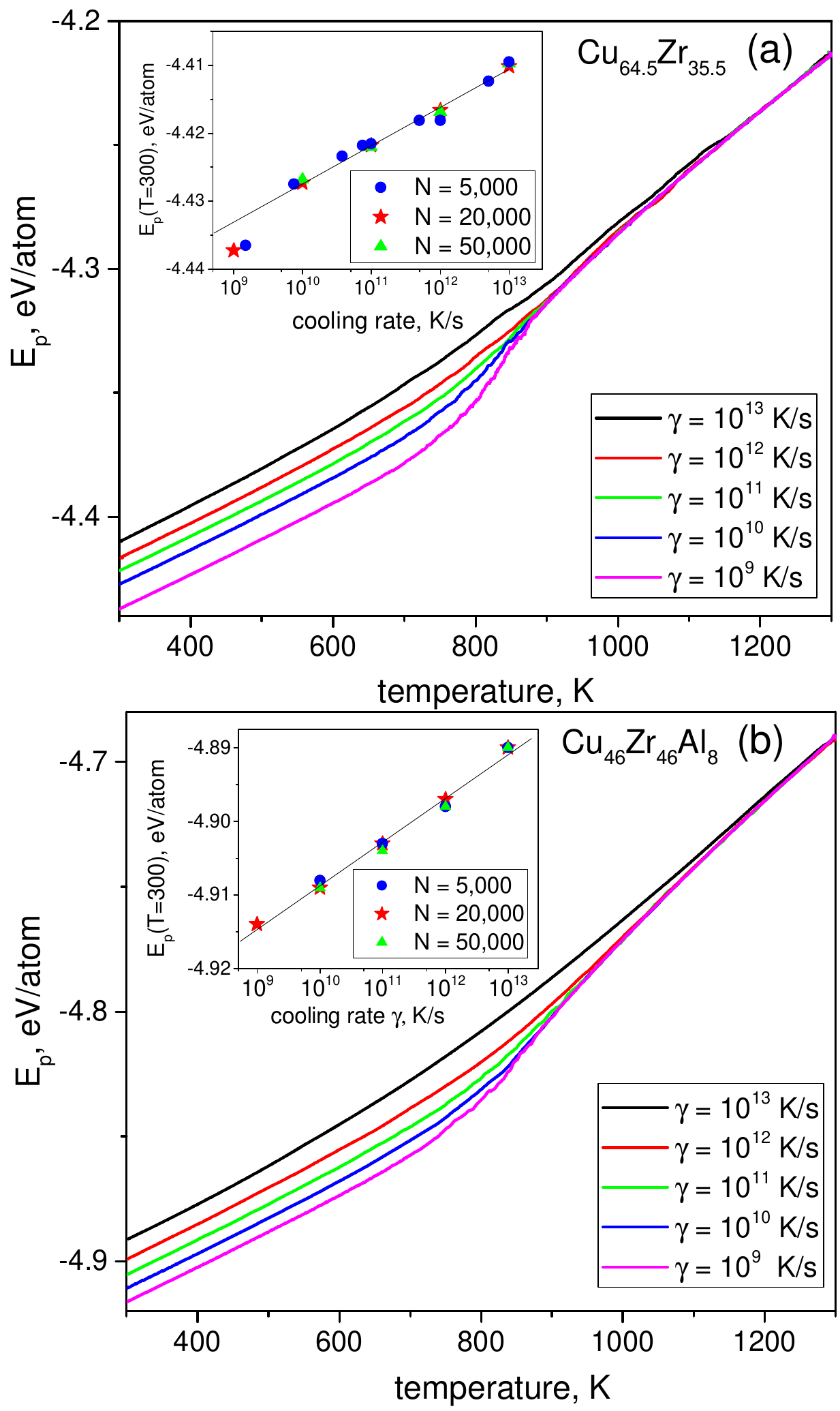}\\
  \caption{ Temperature dependencies of potential energy $E_p$ for (a) ${\rm Cu_{64.5}Zr_{35.5}}$ and (b) ${\rm Cu_{46}Zr_{46}Al_{8}}$ glass-forming alloys at different cooling rates $\gamma$. In both figures, short-time fluctuations of $E_p$ have been smoothed by averaging over appropriate time window. Insets show cooling rate dependencies of room temperature value of potential energy $E^{\rm(300)}_{p}(\gamma)$ obtained at different numbers of particles. For ${\rm Cu_{64.5}Zr_{35.5}}$ alloy, deviation of this dependence from logarithmic law at $\gamma=10^{9}$ K/s is caused by partial crystallization.}
  \label{fig:Ept}
\end{figure}

\begin{figure}
  \centering
  \includegraphics[width=\columnwidth]{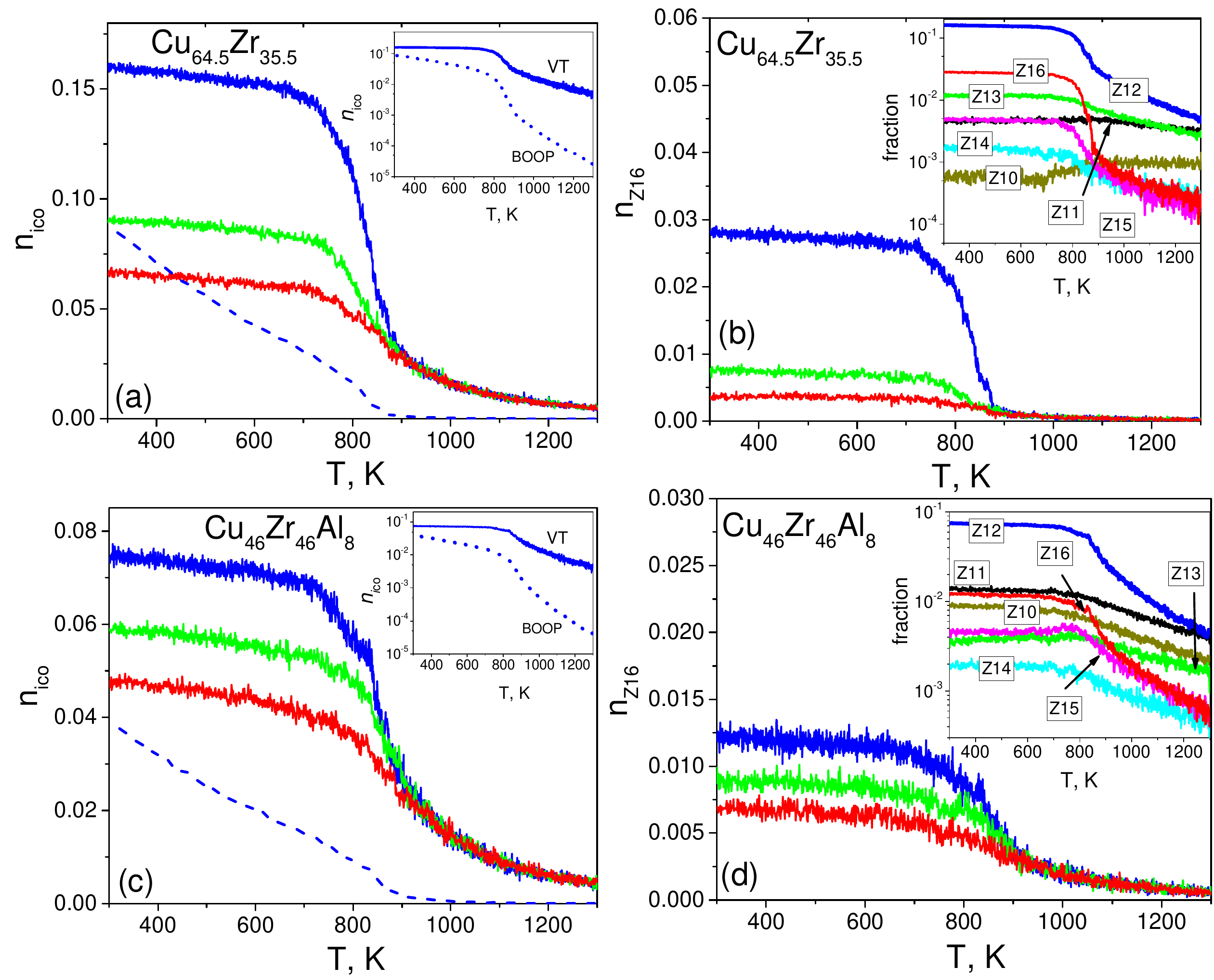}\\
  \caption{ Temperature dependencies of fraction of both icosahedral and Z16 Kasper polyhedra for ${\rm Cu_{64.5}Zr_{35.5}}$ (a,b)  and  ${\rm Cu_{46}Zr_{46}Al_{8}}$ (c,d) glass-forming alloys at different cooling rates $\gamma$. Number of particles in both cases is $N=20,000$. Solid lines represent results obtained by Voronoi analysis and dashed lines are the fractions of icosahedra obtained for $\gamma=10^9$ K/s by BOOP method.  We see that both $n_{\rm ico}$ and $n_{\rm Z16}$ for ${\rm Cu_{64.5}Zr_{35.5}}$ alloy cooled at $\gamma=10^9$ reveal drastic increase at $T \simeq 850$ K caused by crystallization. The insets in panels (a,c) demonstrate difference between fractions of icosahedra obtained by VT and BOOP in log-scale. Insets in panels (b,d) show temperature dependence of abundances of Kasper polyhedra Z10-Z16 calculated at cooling rate of $10^{9} $ K/s.}
  \label{fig:ico}
\end{figure}

 \begin{figure}
  \centering
  \includegraphics[width=\columnwidth]{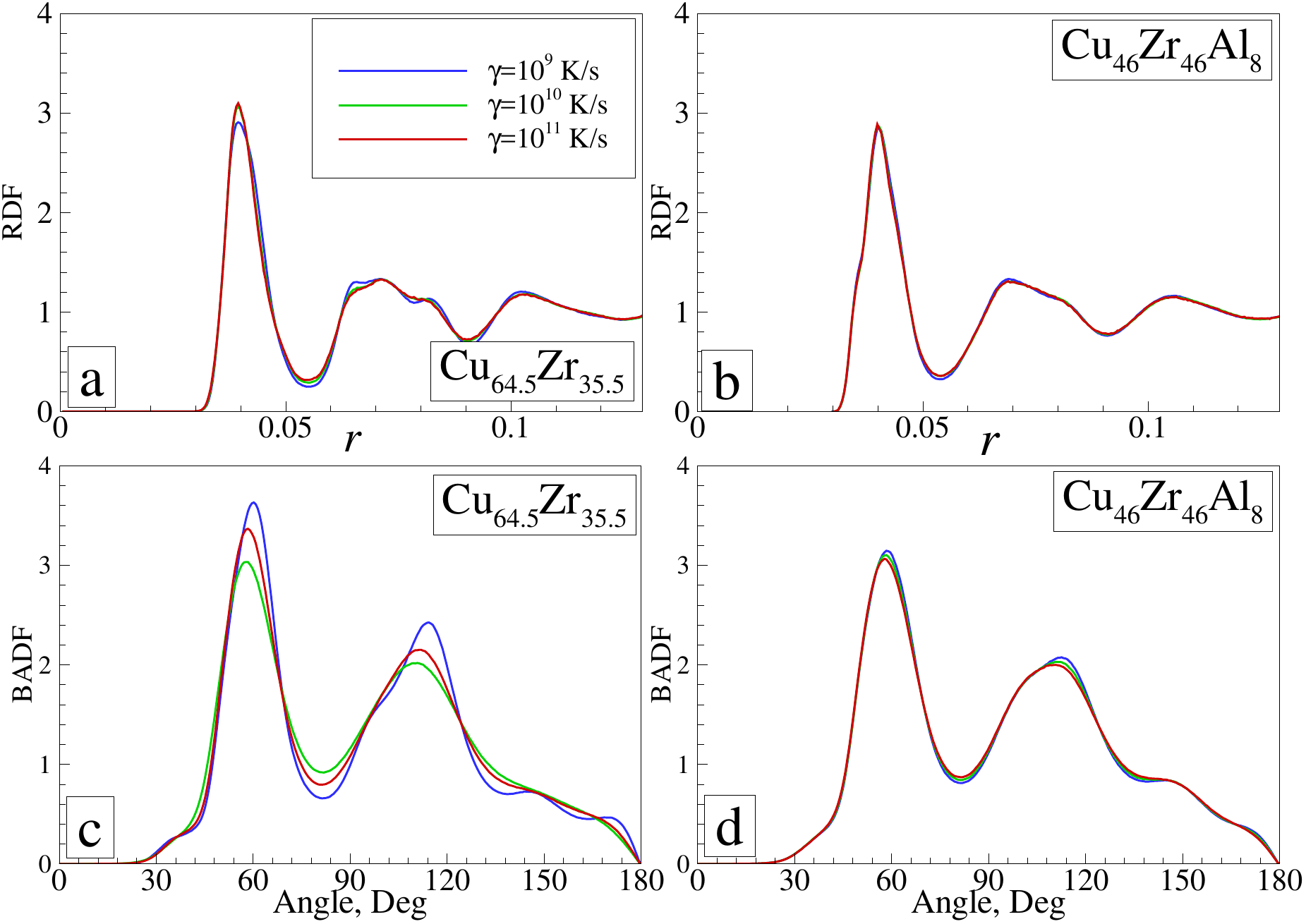}\\
  \caption{Pair correlation functions (RDF and BADF) for final room-temperature states of ${\rm Cu_{64.5}Zr_{35.5}}$ (a,c) and ${\rm Cu_{46}Zr_{46}Al_{8}}$ (b,d) alloys at different cooling rates $\gamma$. The picture suggests that ${\rm Cu_{64.5}Zr_{35.5}}$ is partially crystallized at $\gamma=10^9$ K/s but ${\rm Cu_{46}Zr_{46}Al_{8}}$ one is totally amorphous at all cooling rates.}
  \label{fig:corr}
\end{figure}

For simulated  ${\rm Cu_{64.5}Zr_{35.5}}$ alloy, we have already reported that relatively small $N=5,000$ system crystalizes partially being cooled at $\gamma = 1.5\cdot 10^{9}$~K/s~\cite{Ryltsev2016JCP}. The structure of this sample was a glassy matrix with nano-sized crystalline inclusion of ${\rm Cu_2Zr}$ (C15 Laves phase) compound. The natural question arises how this pictures depends on both system size and cooling rate. Here we address this question for ${\rm Cu_{64.5}Zr_{35.5}}$  mixture and compare the results with ${\rm Cu_{46}Zr_{46}Al_{8}}$ one which has not yet been studied in that context. First we consider structural evolution of the systems under continuous cooling.

In Fig.~\ref{fig:Ept} we show smoothed temperature dependencies of the potential energy $E_{p}$ of the systems with $N=20,000$ particles obtained under continuous cooling at different cooling rates $\gamma$. All the $E_p(T)$ curves demonstrate inflections indicating the liquid-glass transition \cite{Mendelev2009PhilMag_2}. The picture reveals a decrease of the final room temperature value of potential energy $E^{\rm(300)}_{p}$ with decreasing $\gamma$. That means the lower cooling rate is imposed the better system relaxation takes place. The insets in the Fig.~\ref{fig:Ept} show how $E^{\rm(300)}_{p}$ varies with cooling rate $\gamma$ for the systems with different particle number $N$. We see that, for ${\rm Cu_{64.5}Zr_{35.5}}$ alloy, the $E^{\rm(300)}_{p}(\gamma)$ dependencies are well fitted by logarithmic function at $\gamma \leq 10^{10}$ K/s but essentially deviates from this law at $\gamma = 10^{9}$ K/s. For ${\rm Cu_{46}Zr_{46}Al_{8}}$ alloy, $E^{\rm(300)}_{p}(\gamma)$ keeps to logarithmic dependence at all cooling rates. Note that $E^{\rm(300)}_{p}(\gamma)$ demonstrates no substantial dependence on the number of particles $N$ for both systems.

The idea that potential energy of a particle system depends on cooling rate via logarithmic law has been proposed in~\cite{Zhang2014AppPhysLett,Zhang2015PRB_2}. That seems to be quite natural because similar logarithmic law has been proposed for cooling rate dependence of other glassy state observables, such as, for example, the glass transition temperature ~\cite{Samwer1992PRB}. Thus, for the potential energy, the relation  $E^{\rm(300)}_{p}(\gamma) \sim \log (\gamma)$ is expected for completely glassy systems but it should fail in the case of crystallization. For $N=5,000$ particle ${\rm Cu_{64.5}Zr_{35.5}}$ alloy, we recently observed deviation of $E^{\rm(300)}_{p}(\gamma)$ from logarithmic law caused by partial crystallization\cite{Ryltsev2016JCP}. Here we reveal that for larger system at lower cooling rate this effect is even more pronounced (compare two curves in the inset for Fig.~\ref{fig:Ept}).

\begin{figure}
  \centering
  \includegraphics[width=0.9\columnwidth]{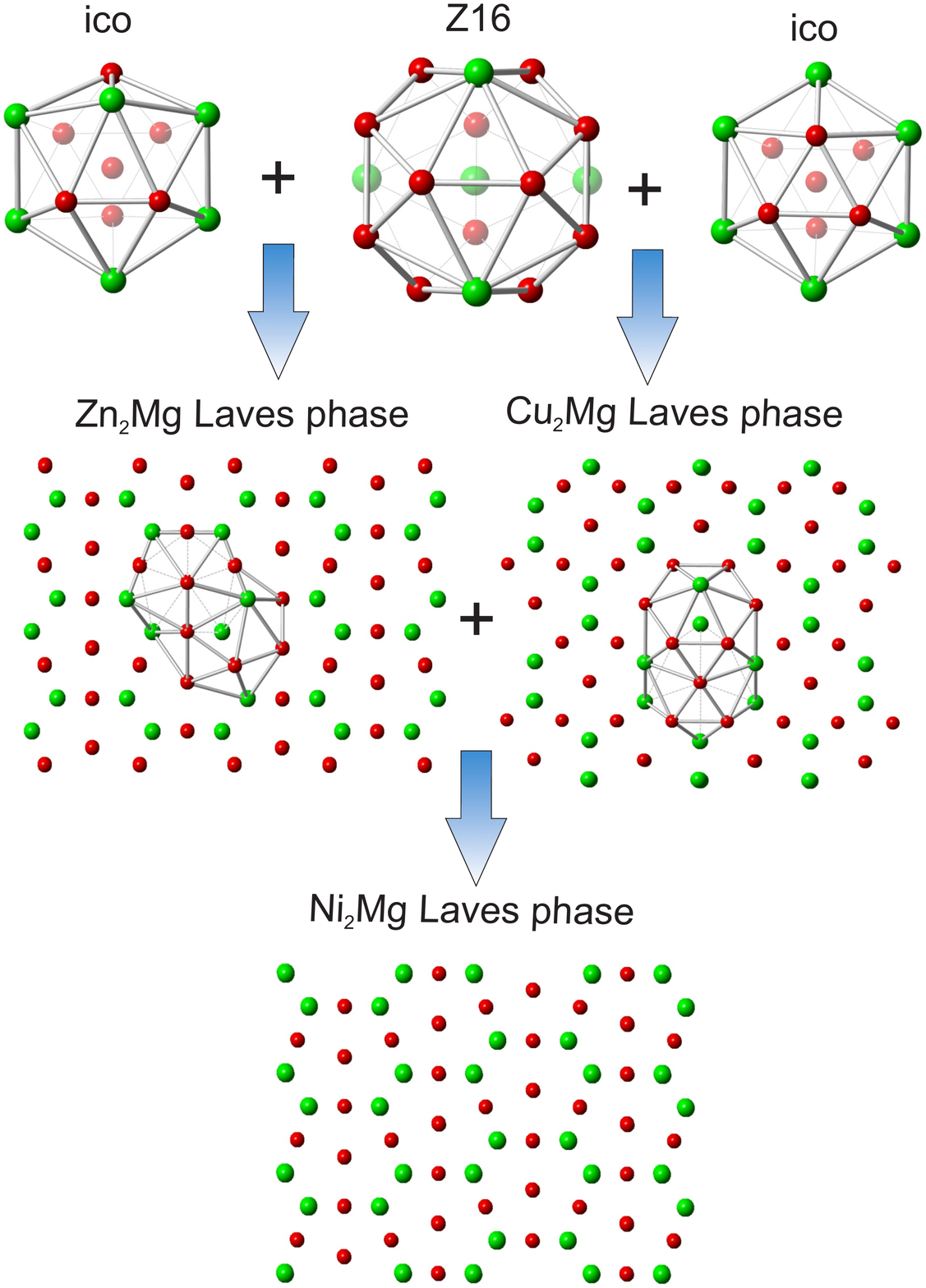}\\
  \caption{We show how three different structures of ${\rm Cu_2Zr}$ compound -- ${\rm Cu_2Mg}$ (C15), ${\rm Zn_2Mg}$ (C14)and ${\rm Ni_2Mg}$ (C36) Laves phases -- can be build from icosahedron and Kasper polyhedron Z16. Icosahedra for C15 and C14 structures have different distribution of Cu and Zr atoms over the vertexes. Two different joining mechanisms of icosahedra and Z16 are also presented. The structure of C36 Laves phase is locally constructed from both types of local elements tiled properly in the space.}
   \label{fig:laves}
\end{figure}

Described evolution of potential energy under cooling is in close correlation with that for structural characteristics.  Laves phases, including C15 formed under crystallization of ${\rm Cu_{64.5}Zr_{35.5}}$ alloy, are build of both icosahedra and Kasper polyhedra Z16 (see Fig.~\ref{fig:laves} and explanation in the text). So the fraction of these clusters can serve as indicators of crystallization in the case of Laves phase formation. In Fig.~\ref{fig:ico}, we show temperature dependencies of fractions of particles, which are the centers of icosahedral-like clusters ($n_{\rm ico}$) and Z16 ones ($n_{\rm Z16}$), calculated for the systems under consideration at different cooling rates. The clusters were mainly identified by VT and BOOP method was applied for comparison.  We see from the picture that both $n_{\rm ico}$ and $n_{\rm Z16}$ increase with decreasing temperature and demonstrate drastic growth at the temperatures where inflections on $E_p(T)$ curves occur. That means drastic change of the local structure at the glass transition temperature. Note that the kinks of $n_{\rm ico}(T)$ and $n_{\rm Z16}(T)$ for ${\rm Cu_{64.5}Zr_{35.5}}$ alloy cooled at $\gamma = 10^{9}$ K/s are much more pronounced that is caused by partial crystallization of the system. This effect is particularly evident for the $n_{\rm Z16}(T)$ dependence because Kasper Z16 polyhedra fraction in relatively low in liquid and glassy states but high in Laves phases.

Insets in panels (b,d) of Fig.~\ref{fig:ico} show temperature dependence of fractions of Z10-Z16 Kasper polyhedra (including icosahedra Z12) calculated at cooling rate of $10^{9}$ K/s. We see that icosahedra (Z12) are the most popular local clusters in both systems under consideration. For ${\rm Cu_{64.5}Zr_{35.5}}$ alloy, only the fraction of Z12 and Z16 clusters, which are the building blocks of Laves phases, demonstrate drastic increase under cooling. It also important that both systems demonstrate comparable total amounts of Z10-Z16 polyhedra in liquid and glassy states though partial ZCN fractions are essentially different. That means the difference in structure related properties like GFA (if any) would not be caused by difference in local icosahedral (polytetrahedral to be more general) ordering.

It is interesting, that temperature dependencies of fractions of local clusters obtained by VT and BOOP methods are essentially different (see Fig.~\ref{fig:ico}a,c). First, we see that fractions of icosahedral clusters obtained by VT are much higher than those calculated by BOOP. That is due to the fact that VT determines icosahedra as  $\langle 0, 0, 12, 0 \rangle$ Voronoi polyhedra, which are topologically stable for distortions. So VT-icosahedra may be in fact strongly distorted \cite{Klumov2016JETPLett}. BOOP method, with using sufficiently rigid criteria for cluster determination (like  $q_6 > 0.6$, $w_6 < -0.16$ we use), generates icosahedra with rather high quality. So we can conclude that all the cluster satisfied to the condition  $q_6 > 0.6$, $w_6 < -0.16$ are polyhedra with Voronoi index $\langle 0, 0, 12, 0 \rangle$ but the opposite is not true. This holds for any type of local clusters.

The same peculiarities of VT and BOOP methods also cause difference in $n_{\rm ico}(T)$ dependence at low temperature region. Indeed, VT-curves demonstrate saturation below glass-transition (crystallization) temperature (that is at $T< 850$ K)
 but BOOP-curves reveal intensive growth of icosahedra fraction down to room temperature. The explanation is simple: most of the clusters (icosahedra in our case) form in the vicinity of $T_g$ (or $T_c$) and than, at lower temperatures, the only distortion of these clusters decrease due to decrease of thermal motion. So the VT demonstrates no essential increase of the cluster fractions because it ignores the distortion degree. On the contrary, a decrease of thermal fluctuations leads to more and more VT-icosahedra satisfying the criteria $q_6 > 0.6$, $w_6 < -0.16$ that causes an increase of $n_{\rm ico}(T)$ obtained by BOOP method.

Summarizing the above results, we suggest that VT is more convenient to study temperature (time) dependencies of the cluster abundances but BOOP is preferable when the cluster quality is important (for example, to study spatial distribution of crystal nuclei).

Thus, analysing Fig.~\ref{fig:Ept},Fig.~\ref{fig:ico}, we argue that ${\rm Cu_{46}Zr_{46}Al_{8}}$ alloy is completely amorphous at all cooling rates used  but ${\rm Cu_{64.5}Zr_{35.5}}$ one crystalizes partially at $\gamma = 10^{9}$ K/s.  To check that we analyse the structure of final room-temperature states for both systems. In Fig.\ref{fig:corr}, we show both RDFs and bong angle distribution functions (BADFs) for such states at different $\gamma$. The picture supports the above suggestion. In particular, for ${\rm Cu_{64.5}Zr_{35.5}}$ cooled at $\gamma=10^9$ K/s,  BADF reveals small peak near $\theta=180^{\circ}$ which indicates the presence of crystalline phase.

\begin{figure}
  \centering
  \includegraphics[width=0.8\columnwidth]{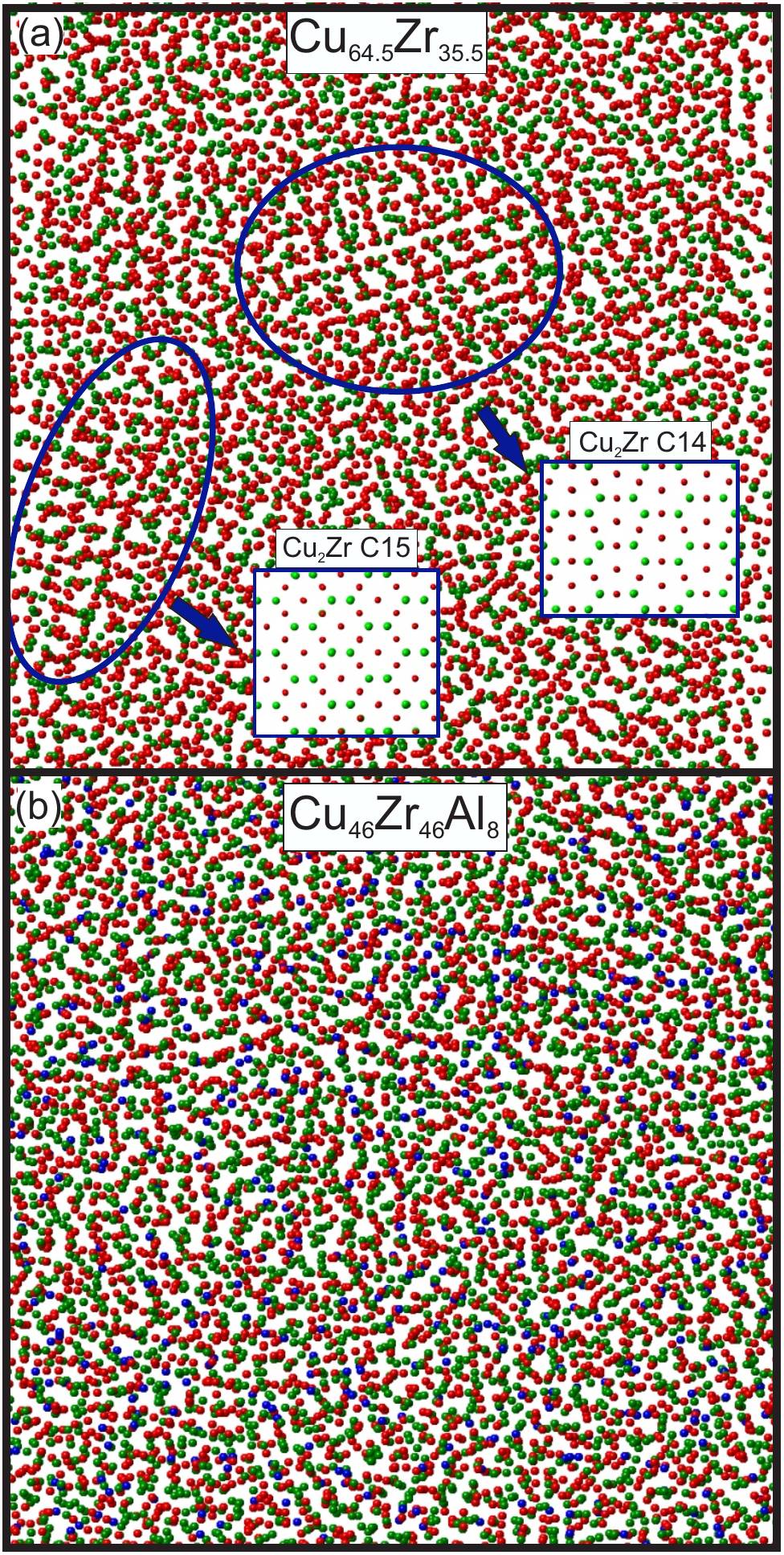}\\
  \caption{ Snapshots of ${\rm Cu_{64.5}Zr_{35.5}}$ (a) and ${\rm Cu_{46}Zr_{46}Al_{8}}$ (b) glass-forming alloys for the final room-temperature states obtained at cooling rate of $10^9$ K/s. Cu, Zr and Al atoms are colored red, green and blue, respectively. The blue ovals indicate crystalline grains of ${\rm Cu_2Zr}$ compounds with structure of ${\rm Cu_2Mg}$ (C15) and ${\rm Zn_2Mg}$ (C14) Laves phases. Both phases have similar local structure consisting of icosahedra and Z16 Kasper polyhedra (see Fig.~\ref{fig:laves}). }
   \label{fig:snapshots}
\end{figure}

To visualise the result mentioned, we present in Fig.~\ref{fig:snapshots} typical snapshots for $\gamma = 10^{9}$ K/s which reveal pronounced ordered domains for ${\rm Cu_{64.5}Zr_{35.5}}$ alloy but visually amorphous structure for the  ${\rm Cu_{46}Zr_{46}Al_{8}}$ one.

As already mentioned, for $N=5,000$ particle ${\rm Cu_{64.5}Zr_{35.5}}$ alloy, the nanocrystal grain formed at rapid cooling is ${\rm Cu_2Zr}$ compound with the structure of C15 Laves phase \cite{Ryltsev2016JCP}. Surprisingly, for $N=20,000$ particle system, we observe two polymorphous modifications of this compound with structures of  C15 and C14 Laves phases (see insets in Fig.~\ref{fig:snapshots}a). These structures are constructed from the same building blocks, such as icosahedron and Kasper polyhedron Z16 (see Fig.~\ref{fig:laves}). At the local order level, the differences between two phases are different distributions of Cu and Zr atoms over the vertexes of icosahedra and different joining mechanism of icosahedral and Z16 clusters  (Fig.~\ref{fig:laves}). Such phases are expected to have close values of free energy and so their coexistence in the same sample is quite understandable. Remember, that the same situation takes place for Lennard-Jones system where hcp and fcc lattices having close values of free energy always coexist in the crystalline phases formed by crystallization of the liquid.

\begin{figure}
  \centering
  \includegraphics[width=\columnwidth]{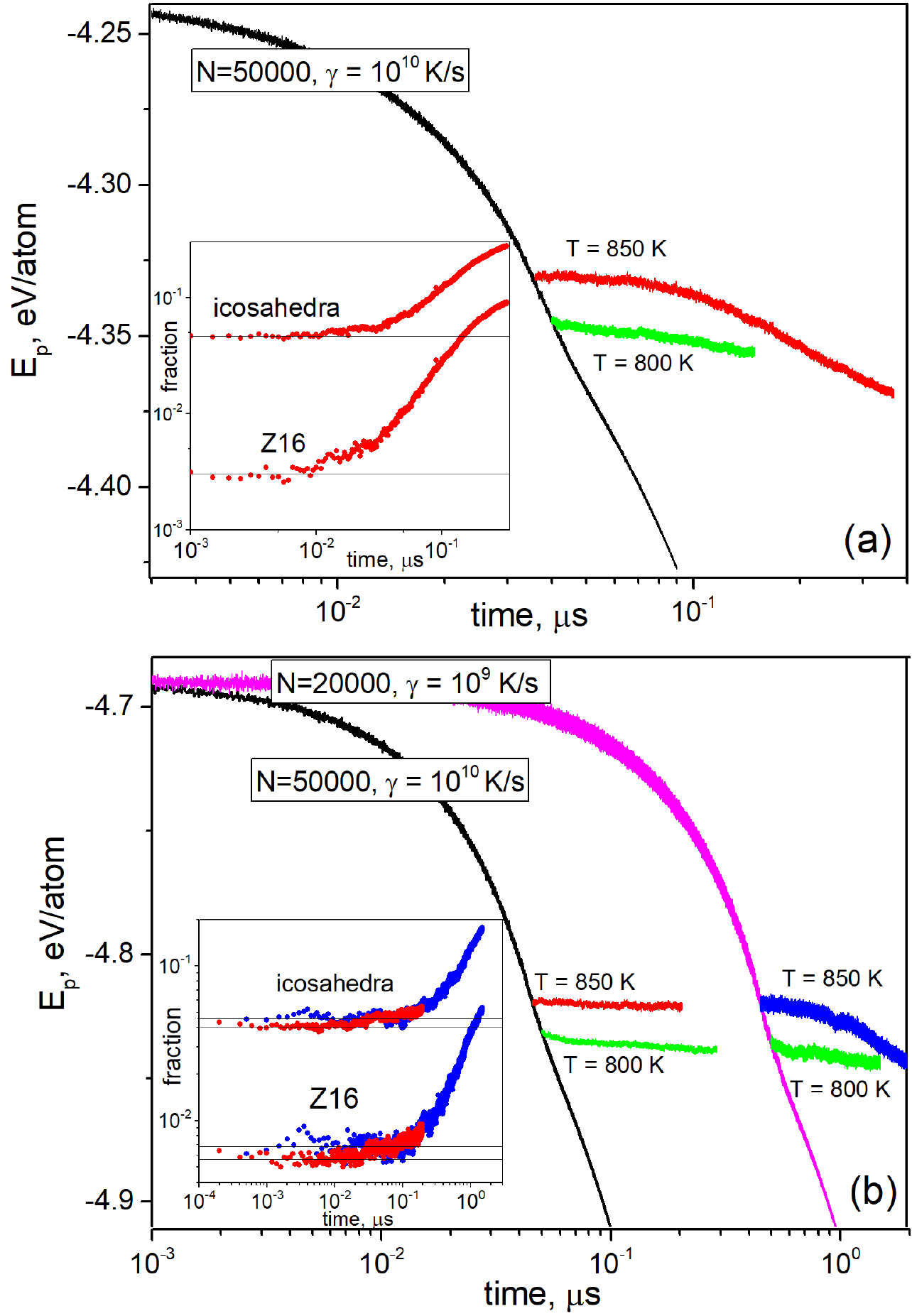}\\
  \caption{Time dependencies of the potential energy during both continuous cooling and isothermal annealing at different temperatures. Cooling rates, numbers of particles and annealing temperatures are indicated on the curves. Insets show time dependencies of fractions of both icosahedral and Z16 clusters, which confirm that the decrease of average $E_p$ during the annealing is caused by crystallization. Red and blue dots in the insets correspond to $T=850$ K annealing at $N=50,000$ and $N=20,000$, respectively}
  \label{fig:ann}
\end{figure}

 \begin{figure}
  \centering
  \includegraphics[width=0.8\columnwidth]{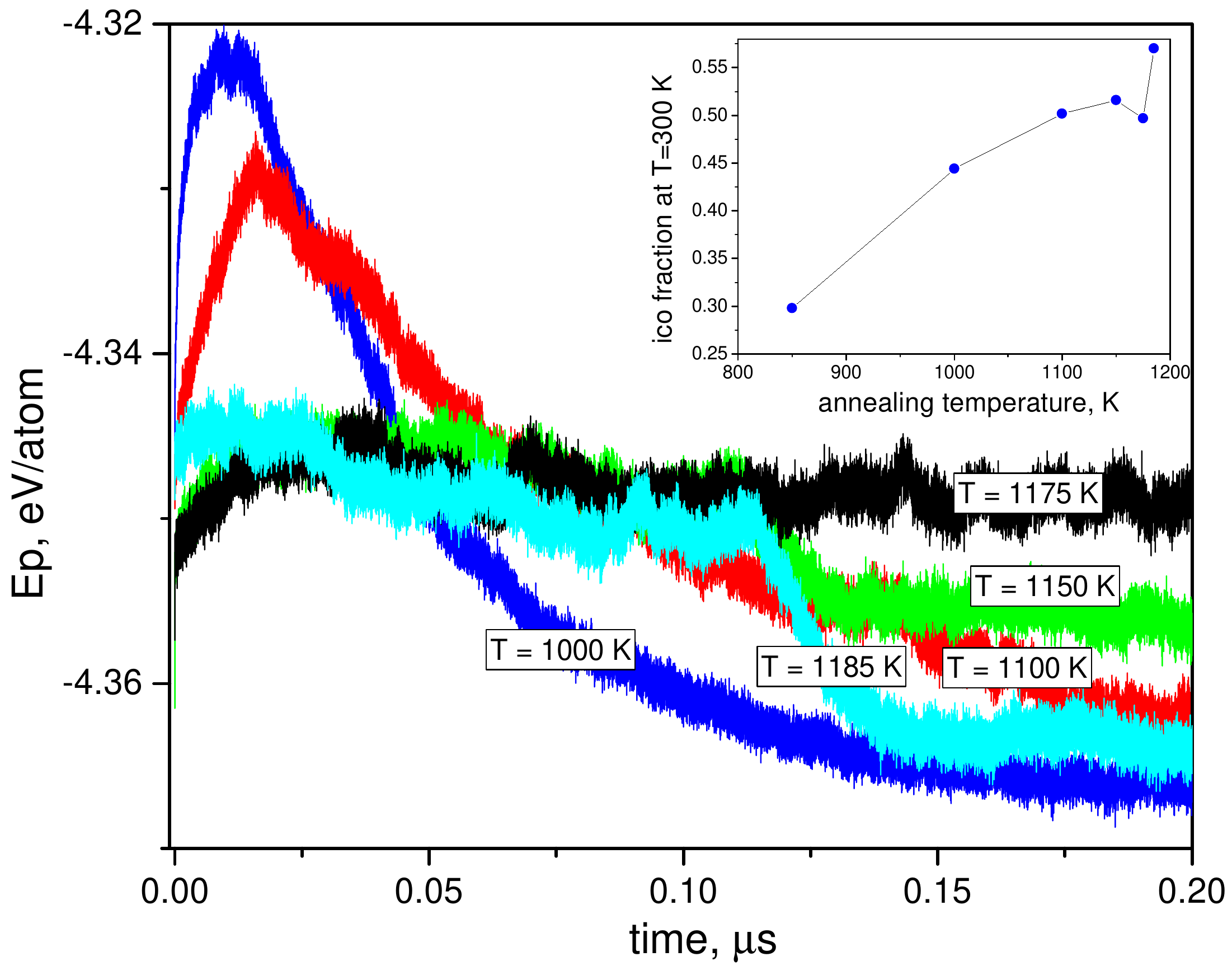}\\
  \caption{Time dependencies of potential energy during five consecutive annealings ${\rm Cu_{64.5}Zr_{35.5}}$ alloy performed after the annealing at $T = 850$ K. The final configuration from previous annealing was taken as the initial configuration for the next one. The curves demonstrate initial increase of $E_p$ caused by an increase of temperature following by a decrease due to the further relaxation (crystal growth and/or recrystallization). Inset demonstrate dependence of fraction of icosahedral clusters on annealing temperature obtained for final configurations which were instantly quenched down to $T=300$ K and then relaxed.}
   \label{fig:ann2}
\end{figure}

\begin{figure}
  \centering
  \includegraphics[width=0.8\columnwidth]{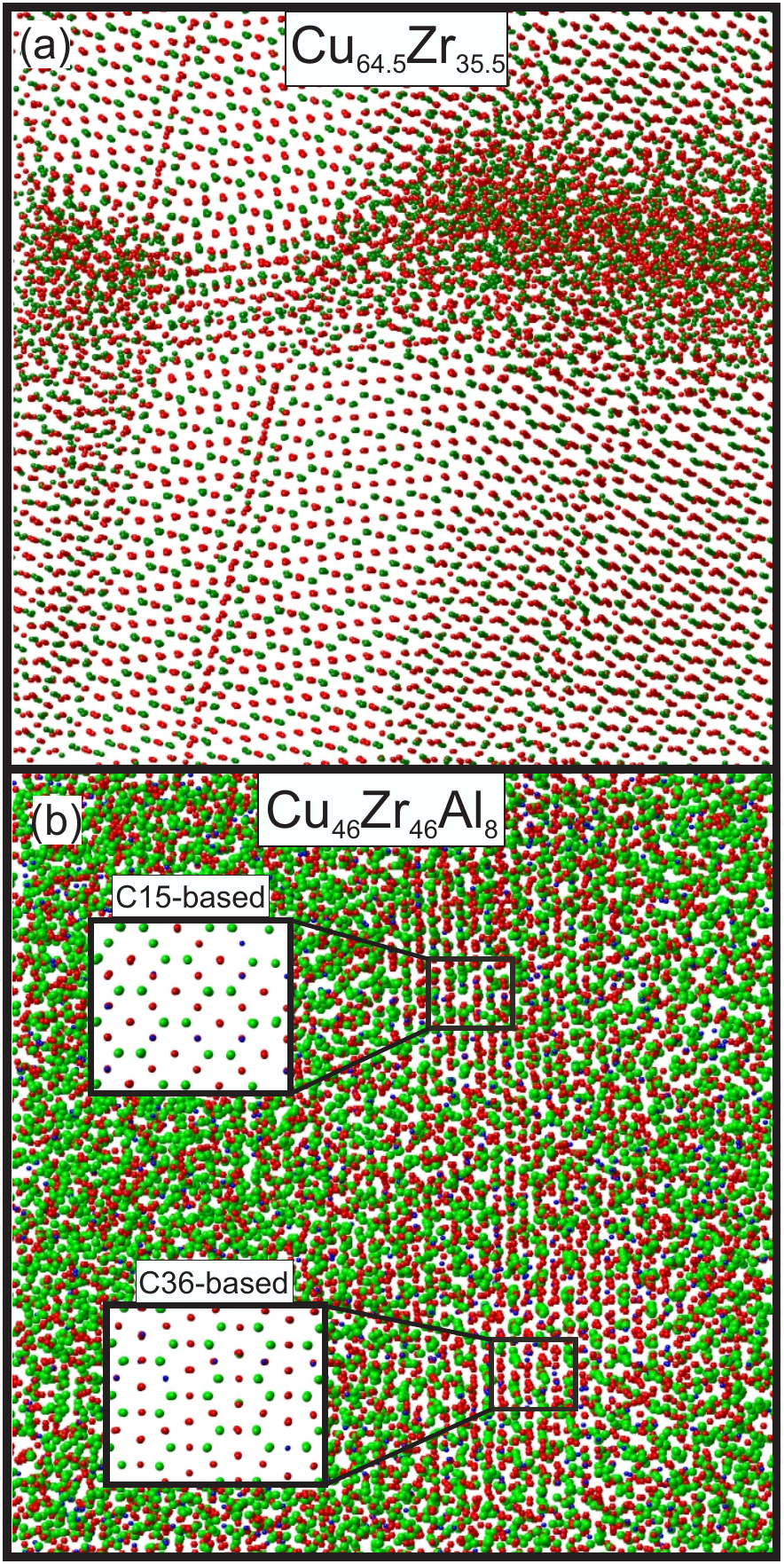}\\
  \caption{Snapshots of the $N=50,000$ particle ${\rm Cu_{64.5}Zr_{35.5}}$ (a) and $N=20,000$ particle ${\rm Cu_{46}Zr_{46}Al_{8}}$ (b) alloys after isothermal annealing. Cu, Zr and Al atoms are colored red, green and blue, respectively. ${\rm Cu_{64.5}Zr_{35.5}}$ alloy was annealed at $T=850$ K for 0.5 $\mu$s and then at $T= (1, 1.1, 1.15, 1.175, 1.185)\cdot 10^3$ K for 0.2 $\mu$s (see Fig.~\ref{fig:ann}a). Resulting structure consists of a few grains of ${\rm Cu_2Zr}$ intermetallic compound with structure of ${\rm Cu_2Mg}$ (C15) Laves phase; small amounts of ${\rm Zn_2Mg}$ (C14) phase as well as disordered glassy-like phase are also presented. ${\rm Cu_{46}Zr_{46}Al_{8}}$ alloy was annealed at $T=850$ K for 3 $\mu$s (see Fig.~\ref{fig:ann}a) and then at $T=1000$ K for 0.5 $\mu$s. The resulting structure is the mixture of amorphous phase and a few crystalline grains. These grains are solid solutions based on ${\rm Cu_2Zr}$ compound with structure of both ${\rm Cu_2Mg}$ (C15) and ${\rm Ni_2Mg}$ (C36) Laves phases (see insets in panel (b)).}
   \label{fig:snapshots_ann}
\end{figure}

\subsection{Structural evolution during isothermal annealing}

Previously, we demonstrated that isothermal annealing of ${\rm Cu_{64.5}Zr_{35.5}}$ $N=5,000$ particle system near the glass-transition temperature leads to the formation of the same ${\rm Cu_2Zr}$ nucleus as for continuous cooling~\cite{Ryltsev2016JCP}. Here we do the same annealing for the case of $N=50,000$ for both ${\rm Cu_{64.5}Zr_{35.5}}$ and ${\rm Cu_{46}Zr_{46}Al_{8}}$ alloys (see Fig.~\ref{fig:ann}). Initial configurations for annealing were collected from those obtained under continuous cooling at $\gamma = 10^{10}$ K/s. We see from the Fig.~\ref{fig:ann}a that averaged value of potential energy for ${\rm Cu_{64.5}Zr_{35.5}}$ alloy decreases during the annealing that means system relaxation. The most pronounced relaxation is observed at $T=850$ K. Analysis of structure evolution reveals that the decrease of $E_p$ is caused by crystallization. To demonstrate that, we show in the insets of Fig.~\ref{fig:ann} time dependencies of fraction of icosahedral and Z16 clusters. Monotonous increase of these fractions $t > 0.01$ $\mu$s reveals nucleation and growth of Laves phases.

As follows from the Fig.~\ref{fig:ann}b, ${\rm Cu_{46}Zr_{46}Al_{8}}$ alloy does not demonstrate essential relaxation at annealing of $N=50,000$ particle system at least for 0.2 $\mu$s. However, time dependencies of $n_{\rm ico}$ and $n_{\rm Z16}$ demonstrate slight increase at $t> 0.3 $ $\mu$s, which can be caused by initial growth of crystal nuclei. Simulation of $N=50,000$ at such timescales requires about a month of calculations and so further relaxation would be too time consuming. Instead, we perform annealing of $N=20,000$ particle system that allows us to extend simulation timescale up to 1.5 $\mu$s ($7.5\cdot 10^8$ MD step). As the result of such long annealing at $T=850$ K, we observe essential decrease of average potential energy (see, Fig.~\ref{fig:ann}b). An increase of $n_{\rm ico}(t)$ and $n_{\rm Z16}(t)$ during the annealing (see, inset in Fig.~\ref{fig:ann}b) suggests that this relaxation is caused by nucleation and growth of Laves phases. The analysis of the structure confirm this idea (see below).

 Isothermal annealing of a particle system at any temperature, which is less that crystallization temperature (solidus or liquidus in the case of mixtures), must eventually lead to the formation of an equilibrium amount of crystalline phase, which is determined by phase diagram of the system (stable or metastable). For the systems under consideration, binodals determining the equilibrium between the liquid mixture and the ${\rm Cu_2Zr}$ compound (${\rm Cu_2Zr}$-based solid solution for the ternary system) are not available. However, it follows from the general reasons that a fraction of a crystal phase formed during an isothermal annealing in a finite time is a result of competition between thermodynamic and kinetic factors. Moreover, annealing of polycrystalline configurations may cause recrystallization due to relaxation of the system towards monocrystalline state (for example, growth of the crystalline grains and their reorientation).  So the question arises which annealing temperature $T_{\rm ann}$ is optimal to obtain better structure with highest fraction of crystalline phase (Laves phase in our case) and lowest density of structural defects (grain boundaries). To address this issue, we consider ${\rm Cu_{64.5}Zr_{35.5}}$ alloy, partially crystallized after the $T=850$ K annealing, and perform the series of its consecutive isothermal annealings at $T= (1, 1.1, 1.15, 1.175, 1.185)\cdot 10^3$ K taking the final configuration from previous MD run as the initial configuration for the next one. Note that the melting (liquidus to be more exact) temperature of ${\rm Cu_2 Zr}$ compound is about $T_m\approx 1230$ K that means we use annealing at $T/T_m \in (0.81, 0.96)$.  The time dependencies of potential energy $E_p(t)$ for such annealings demonstrate initial increase of $E_p$ caused by an increase of temperature following by a decrease due to the further relaxation (crystal growth and/or recrystallization) (see Fig.~\ref{fig:ann2}). Analysis of the snapshots reveals more and more pronounced crystalline structure after each annealing. Note that, for the annealing at $T=1185$ K, we observe drastic drop of  $E_p$ at $t\simeq 0.12$ $\mu$s, which was found to be caused by recrystallization accompanied by growth of crystalline grains of ${\rm Cu_2Mg}$ (C15) phase and disappearance of ${\rm Zn_2Mg}$ (C14) ones. Such behavior suggests that C15 phase in stable for the ${\rm Cu_{64.5}Zr_{35.5}}$ alloy but C14 is metastable one.

  To compare quantitatively the structure of the final configurations for each annealing, we perform their rapid quench down to $T=300$ K and subsequent relaxation at such temperature. The inset in  Fig.~\ref{fig:ann2} shows a dependence of icosahedra fraction for such states on annealing temperature $T_{\rm ann}$. At $T_{\rm ann}\leq 1150$ K, we see monotonous growth of icosahedra fraction and so amount of crystalline phase; at $T_{\rm ann} \simeq 1175$ K, this fraction starts to decrease and then suddenly rise at $T_{\rm ann} \simeq 1185$ K, that is in agreement with the kink on $E_p(t)$ for this temperature. So, for ${\rm Cu_{64.5}Zr_{35.5}}$ alloy, the maximal fraction of crystal phase is achieved during the annealing at $T=1185$ K.  The final structure after this annealing is the polycrystalline sample consisting of a few highly ordered grains of ${\rm Cu_2Zr}$ C15 phase with a small portion of C14 phase (Fig.\ref{fig:snapshots_ann}).

  For ${\rm Cu_{46}Zr_{46}Al_{8}}$ alloy, nucleation and growth processes are much slower than for ${\rm Cu_{64.5}Zr_{35.5}}$ one and so we do not perform the same multistage annealing for the former. Hoverer, we perform additional one-step annealing of the alloy at $T=1000~K\simeq 0.9T_m$, which also leads to further relaxation and allows obtaining better crystalline structure.  Analysis of the final structure reveals a few nanocrystalline grains embedded into disordered glassy matrix  (Fig.~\ref{fig:snapshots_ann})b). These grains are substitutional solid solutions based on binary ${\rm Cu_2Zr}$ compound with the structure of both ${\rm Cu_2Mg}$ (C15) and ${\rm Ni_2Mg}$ (C36)  Laves phases; the Al atoms substitute partially the Cu ones (see insets in Fig.~\ref{fig:snapshots_ann})b).

 The very fact that structure of crystal grains formed in ${\rm Cu_{46}Zr_{46}Al_{8}}$ alloy is similar to that for ${\rm Cu_{64.5}Zr_{35.5}}$ one is quite surprising because the systems under consideration have different composition and their components interact via different interatomic potentials (see discission in sec.~\ref{sec:discuss}). It is also interesting that both ${\rm Cu_{64.5}Zr_{35.5}}$ and ${\rm Cu_{46}Zr_{46}Al_{8}}$ alloys demonstrate similar polymorphism of ${\rm Cu_2Zr}$ compound: C15/C14 and C15/C36, respectively.

\subsection{Impact of the system size on crystallization process}

 \begin{figure}
  \centering
  \includegraphics[width=\columnwidth]{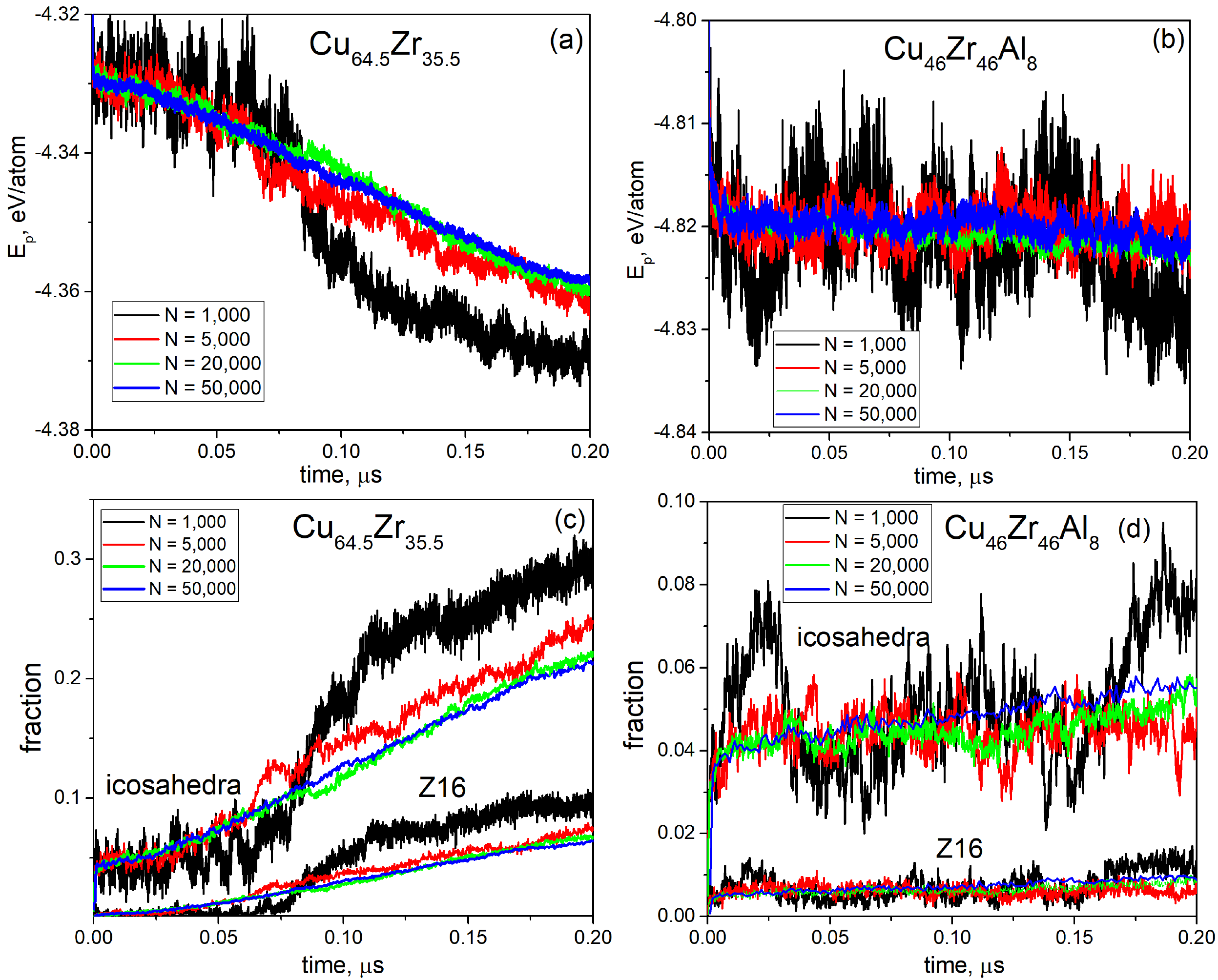}\\
  \caption{Time dependencies of potential energy (a,b) and fractions of icosahedral and Z16 clusters (c,d) during isothermal annealing of ${\rm Cu_{64.5}Zr_{35.5}}$ and ${\rm Cu_{46}Zr_{46}Al_{8}}$ super-cooled liquids of different sizes. Initial states are melts equilibrated at $T=1300$ K and then quenched immediately down to annealing temperature $T=850$ K.}
   \label{fig:ann3}
\end{figure}

The results presented above raise a question about an impact of the system size on the processes of nucleation and growth in the systems under consideration. First, cooling the systems with different cooling rates at different number of particles $N$ reveals no essential size effects at least in the range of $N\in (5,000 -50,000)$ used (see insets in Fig.~\ref{fig:Ept}). Second, for the annealing of ${\rm Cu_{46}Zr_{46}Al_{8}}$ alloy, we see that characteristic nucleation time scale as well as the rate of the growth of the crystal phase are similar for systems of different sizes (Fig.~\ref{fig:ann}b).

These results are rather interesting in the context of finite-size effects in simulations of supercooled liquids and their nucleation stability. This issue has long-standing history, from early simulations of a few hundreds particle systems to recent simulations of millions particle ones \cite{Mandell1977JCP,Honeycutt1986JPC,Swope1990PRB,Streitz2006PRB}. According to general considerations, the critical nucleation time should demonstrate $1/N$ dependence on system size \cite{Honeycutt1986JPC}. Thus, the larger a system is, the faster it has to crystalize.  However, for small $N \lessapprox 1,000$ systems this rule may fail due to disturbance of nuclei formation by periodic boundary condition \cite{Honeycutt1986JPC}.

The comparison of $N$ dependence of critical nucleation time on the basis of the results considered above is rather ambiguous because the systems with different sizes were in different initial conditions. For example, as seen from the inset in Fig.~\ref{fig:ann}b, the critical nucleation time for ${\rm Cu_{46}Zr_{46}Al_{8}}$ alloy is slightly less at $N=50,000$ than at $N=20,000$. However, the initial state for latter system was obtained at lower cooling rate and so the initial fraction of icosahedral and Z16 clusters was also different.

 To eliminate possible impact of initial state and study size effects more consistently, we perform isothermal annealing of both ${\rm Cu_{64.5}Zr_{35.5}}$ and ${\rm Cu_{46}Zr_{46}Al_{8}}$ alloys at different system sizes $N=(1, 5, 20, 50 )\cdot 10^3$.  In all cases, the initial configurations were high-temperature melts equilibrated at $T=1300$ K and than quenched immediately down to annealing temperature $T=850$ K. In Fig.~\ref{fig:ann3} we show time dependencies of potential energy as well as fractions of both icosahedral and Z16 clusters during such annealing. Formation of critical crystalline nuclei is a random process an so rigorous calculation of critical nucleation $t_{n}$  requires averaging over several independent MD runs \cite{Ingebrigtsen2018crystallisation}. However, as a rough estimation, we can obtain  $t_{n}$ from the pictures as the time where the growth of the clusters (or decrease of $E_p$) starts. Doing so we conclude that $t_{n}$ in ${\rm Cu_{64.5}Zr_{35.5}}$ alloy decreases with increasing $N$ (see Fig.~\ref{fig:ann3}a,c). Indeed,  $t_n\simeq 0.07$ $\mu$s for $N=1,000$  and $t_n\simeq 0.03$ $\mu$s for $N=5,000$. For $N\geq 20,000$,  $t_{n}$ in ${\rm Cu_{64.5}Zr_{35.5}}$ alloy is vanishingly small and it is hardly detectable from the picture. For ${\rm Cu_{46}Zr_{46}Al_{8}}$ alloy, such estimations are even more hardly obtainable at the simulation timescales presented in Fig.~\ref{fig:ann3}b,d. We see that, at $N \leq 5,000$, no essential crystal growth is detectable for $t\leq 0.2$ $\mu$s. At $N = 20,000$, $t_n\simeq 0.12$ $\mu$s and, at $N = 50,000$, $t_n\simeq 0.08$. So the decrease of critical nucleation time with increasing $N$ is observed as well.

 Above results enable to make two important conclusions. First, simulated ${\rm Cu_{64.5}Zr_{35.5}}$ alloy described by EAM potential of Mendelev et.al. is inherently unstable to crystallization at large system sizes and so it is in fact very poor glassformer. Second, the critical nucleation time for ${\rm Cu_{46}Zr_{46}Al_{8}}$ alloy is of the order of magnitude higher than that for ${\rm Cu_{64.5}Zr_{35.5}}$ one that means the former is much more stable to crystallization than the latter. The reasons for such behaviour will be discussed below (see section~\ref{sec:discuss}).

Recently, Ingebrigtsen et al. uncovered new mechanism of crystallisation in multicomponent systems related to compositional fluctuations in the supercooled liquids \cite{Ingebrigtsen2018crystallisation}. The growth of such fluctuations at an increase of system size leads to regions of one species which are larger than the critical nucleus size for crystallisation of that species only. That means multicomponent mixtures are inherently unstable to crystallisation at large enough system sizes. The authors show that, due to this mechanism, the Kob-Andersen model is in fact very poor glassformer at $N>10,000$. On the first glance, compositional fluctuations must "kill" GFA of any mixture those individual components are not good glassformers. However, as was shown in \cite{Ingebrigtsen2018crystallisation}, the mean size of compositional fluctuations $\langle n_s(N)\rangle$  demonstrates slow logarithmic growth with $N$ and so different systems achieve the system size where $\langle n_s(N)\rangle$ is of the order of the critical nucleus size at essentially different numbers of particles. For Kob-Andersen model this "critical" system size is $N_c \sim 10^4$ but for EAM model of ${\rm Cu_{64.5}Zr_{35.5}}$ alloy $N_c \sim 10^{16}$. That means crystallization stability of  ${\rm Cu_{64.5}Zr_{35.5}}$ alloy (and obviously ${\rm Cu_{46}Zr_{46}Al_{8}}$ too) is not essentially affected by compositional fluctuations at system sizes available for simulations.

\section{Discussion and conclusions\label{sec:discuss}}

We have shown that two widely used EAM models of ${\rm Cu_{64.5}Zr_{35.5}}$ and ${\rm Cu_{46}Zr_{46}Al_{8}}$ glass-forming alloys crystallize in sufficiently lengthly molecular dynamics simulations. Doing computational quenching of both systems at $N=20,000$ and different cooling rates $\gamma\in(10^9,10^{13})$ K/s, we observe that ${\rm Cu_{64.5}Zr_{35.5}}$ alloy crystallizes partially at $\gamma=10^9$ K/s but ${\rm Cu_{46}Zr_{46}Al_{8}}$ is completely amorphous at all $\gamma$. Then we do isothermal annealing of the systems at $N=50,000$ for about 0.5 $\mu$s which leads to almost complete crystallization for ${\rm Cu_{64.5}Zr_{35.5}}$ alloy but still disordered glassy structure for ${\rm Cu_{46}Zr_{46}Al_{8}}$ one. So we conclude that ternary ${\rm Cu_{46}Zr_{46}Al_{8}}$ alloy is much more stable to crystallization than the binary ${\rm Cu_{64.5}Zr_{35.5}}$ one. However, doing longer annealing of ${\rm Cu_{46}Zr_{46}Al_{8}}$ alloy at $N=20,000$, we indeed observe its partial crystallization.

Isothermal annealing of supercooled liquids at different number of particles (Fig.~\ref{fig:ann3}) allows revealing that critical nucleation times $t_n$ for both systems  decreases with increasing the system size. We conclude, that for ${\rm Cu_{64.5}Zr_{35.5}}$ alloy this time becomes  vanishingly small at $N\geq 20,000$ that means the system is inherently unstable to crystallization. For ${\rm Cu_{46}Zr_{46}Al_{8}}$ alloy, $t_n$ is order of magnitude higher that is in agreement with data obtained at continuous cooling.

We observe that both systems studied crystallize with the formation of different polymorphous modifications of ${\rm Cu_2Zr}$ compound with the structure of Laves phases. Indeed,  ${\rm Cu_{64.5}Zr_{35.5}}$  alloy demonstrates the mixture of ${\rm Cu_2Mg}$ (C15) and ${\rm Zn_2Mg}$ (C14) Laves phases; partially crystalline ${\rm Cu_{46}Zr_{46}Al_{8}}$ contains ${\rm Cu_2Zr}$-based solid solution with structure of both  ${\rm Cu_2Mg}$ (C15) and ${\rm Ni_2Mg}$ (C36) Laves phases. All Laves phases can be build from icosahedron and Z16 Kasper polyhedron (see Fig.~\ref{fig:laves}). These structures probably have close values of free energy and so compete with each other during crystallisation.  However, such polymorphous structure should be metastable; at longer annealing times it would relax into mono-phase structure with the only type of Laves phase. For ${\rm Cu_{64.5}Zr_{35.5}}$  alloy, this idea is supported by the result of annealing at $T=1185$ K after which the structure with dominating C15 phase formed (see Fig.~\ref{fig:snapshots_ann}a).

The fact that both alloys studied, despite of different composition and different interatomic potentials, crystalize with the formation of similar structures based on Laves phases suggests nucleation mechanism in the systems under consideration. Laves phases belong to Frank-Kasper family, which are topologically close pack phases build from tetrahedral units. Such structures are locally related with polytetrahedral short-range order observed in both alloys studied. Besides icosahedra and Z16 considered here (see Fig.~\ref{fig:ico}), other  Kasper polyhedra, both topologically perfect and distorted, has been found, see insets in Fig.~\ref{fig:ico}b,d and Refs.~\cite{Li2009PRB,Peng2010ApplPhysLett,Wu2013PRB,Wang2015JPhysChemA,Cheng2009PRL}. These clusters were also found by a genetic algorithm analysis as dominant structural motifs in Cu-Zr alloys \cite{Sun2016SciRep}. At relatively small timescales, polytetrahedral clusters disturb formation of crystalline nuclei and so prevent crystallization. But, at larger timescales, local ordering of both icosahedra and Kasper polyhedra lead to formation of crystalline fragments with Laves phase structure. This mechanism was actually proposed by Frank and Kasper~\cite{Frank1958ActCryst,Frank1959ActCryst} to explain the structure of a broad class of metallic alloys.

Proposed nucleation mechanism allows suggesting the reason why ${\rm Cu_{64.5}Zr_{35.5}}$ alloy described by EAM potential of Mendelev et al. is in fact rather poor glassformer at large system sizes. The composition of the alloy is very close to that for ${\rm Cu_2 Zr}\equiv {\rm Cu_{66.6}Zr_{33.4}}$ compound, which has stable Laves phase structure for this potential. Taking into account pronounced polytetrahedral order in the system, we argue that rather law nucleation threshold for Laves phases takes place. For small system sizes, $N < 20,000$, nuclei formation is disturbed by periodic boundary condition \cite{Honeycutt1986JPC} and so the system demonstrate intermediate GFA. For ${\rm Cu_{46}Zr_{46}Al_{8}}$ alloy described by EAM potential of Cheng et al, such nucleation threshold is probably much higher due to the different ratio between Cu and Zr and the presence of the third component.

\section{Acknowledgments}
Molecular dynamics simulations were supported by the Russian Science Foundation (grant RNF №14-13-00676). Structural analysis was supported by Russian Science Foundation (grant RNF №18-12-00438). We thank Ural Branch of Russian Academy of Sciences for the access to ''Uran'' cluster. This work has been carried out using computing resources of the federal collective usage center Complex for Simulation and Data Processing for Mega-science Facilities at NRC “Kurchatov Institute”, http://ckp.nrcki.ru/.

\bibliography{our_bib_cu_zr}
\end{document}